\documentclass[12pt]{article}
\pdfoutput=1
\usepackage{yfonts}
\usepackage{color}
\usepackage{mhchem}
\usepackage{xcolor}
\usepackage{csquotes}
\usepackage{cite}
\usepackage{hyperref}
\hypersetup{colorlinks=true,linkcolor=red,anchorcolor=black,citecolor=green}
\usepackage[toc,page]{appendix}
\usepackage{amsfonts}
\usepackage{bbold}
\usepackage{textcomp}
\usepackage[DIV13]{typearea}
\usepackage{amsmath, amsthm, amssymb, mathtools,empheq,latexsym,dsfont}
\usepackage{bbm}
\usepackage{slashed, simplewick}
\usepackage[utf8]{inputenc}
\usepackage{graphicx,placeins}
\usepackage{makeidx}
\usepackage[font=small,labelfont=bf]{caption}
\usepackage{nicefrac}
\usepackage{subfigure}
\usepackage{array, bigdelim,multirow,multicol}
\usepackage[integrals]{wasysym}
\usepackage{fancybox}
\usepackage{bm}
\usepackage{float}
\usepackage{rotating}
\usepackage{colortbl}
\usepackage{booktabs}
\usepackage[top=2cm,textwidth=16.6cm,textheight=22.75cm]{geometry}
\usepackage{doi}
\graphicspath{{immagini/}}

\definecolor{Gray}{gray}{0.92}

\newcommand{\be}{\begin{equation}}
\newcommand{\ee}{\end{equation}}
\newcommand{\bea}{\begin{eqnarray}}
\newcommand{\eea}{\end{eqnarray}}

\makeatletter
\renewcommand*{\@fnsymbol}[1]{\ensuremath{\ifcase#1\or *\or  \mathsection\or \ddagger\or
\dagger\or \mathparagraph\or \|\or **\or \dagger\dagger
\or \ddagger\ddagger \else\@ctrerr\fi}}
\makeatother

\makeatletter
\@addtoreset{equation}{section}
\makeatother

\begin{document}
 \unitlength = 1mm

\setlength{\extrarowheight}{0.2 cm}

\title{
\begin{flushright}
\begin{minipage}{0.2\linewidth}
\normalsize
\end{minipage}
\end{flushright}
 {\Large\bf Neutrino Masses
and Higher Degree Siegel Modular Forms}\\[0.2cm]}
\date{}

\author{
Maibam Ricky Devi$^{1}$
\thanks{E-mail: {\tt deviricky@gmail.com}}
\
\\*[10pt]
\centerline{
\begin{minipage}{\linewidth}
\begin{center}
$^1${\small Department of Physics, Gauhati University, Guwahati 781014, Assam, India} \\[2mm]
\end{center}
\end{minipage}}
\\[6mm]}
\maketitle
\thispagestyle{empty}
\centerline{\large\bf Abstract}
\begin{quote}
\indent
In this work, we have analyzed a neutrino model within the distinct framework of modular forms with degree, $ g>1 $ . This offers a more generalized scenario of modular forms which is popularly known as Siegel modular forms. We explore the implications of this special case of automorphic forms for physics beyond the standard model (BSM) within the lepton sector. In our model, we explicitly treat the Yukawa couplings as Siegel modular forms with both degree and level being equivalent to 2. We restrict our modulus parameter for $ \tau_{1}=\tau_{2} $ spanning within the finite modular  $ S_{4}\times Z_{2} $ space. This helps us for a broader understanding of the multiplets at higher degree and simplifies the process of model building of the fermion masses. At the end, we compute the unknown neutrino oscillation parameters and find the optimal values of the modulus parameters $ \tau_{1} $  and $ \tau_{3} $ for which the values of the Yukawa couplings are consistent at $ 3\sigma $ for the input parameters of neutrinos as given in  NuFIT 5.2 and discuss its underlying physics.
\end{quote}
\newpage

\section{Introduction}
\begin{figure}[t!]
\centering
\includegraphics[width=0.5\linewidth]{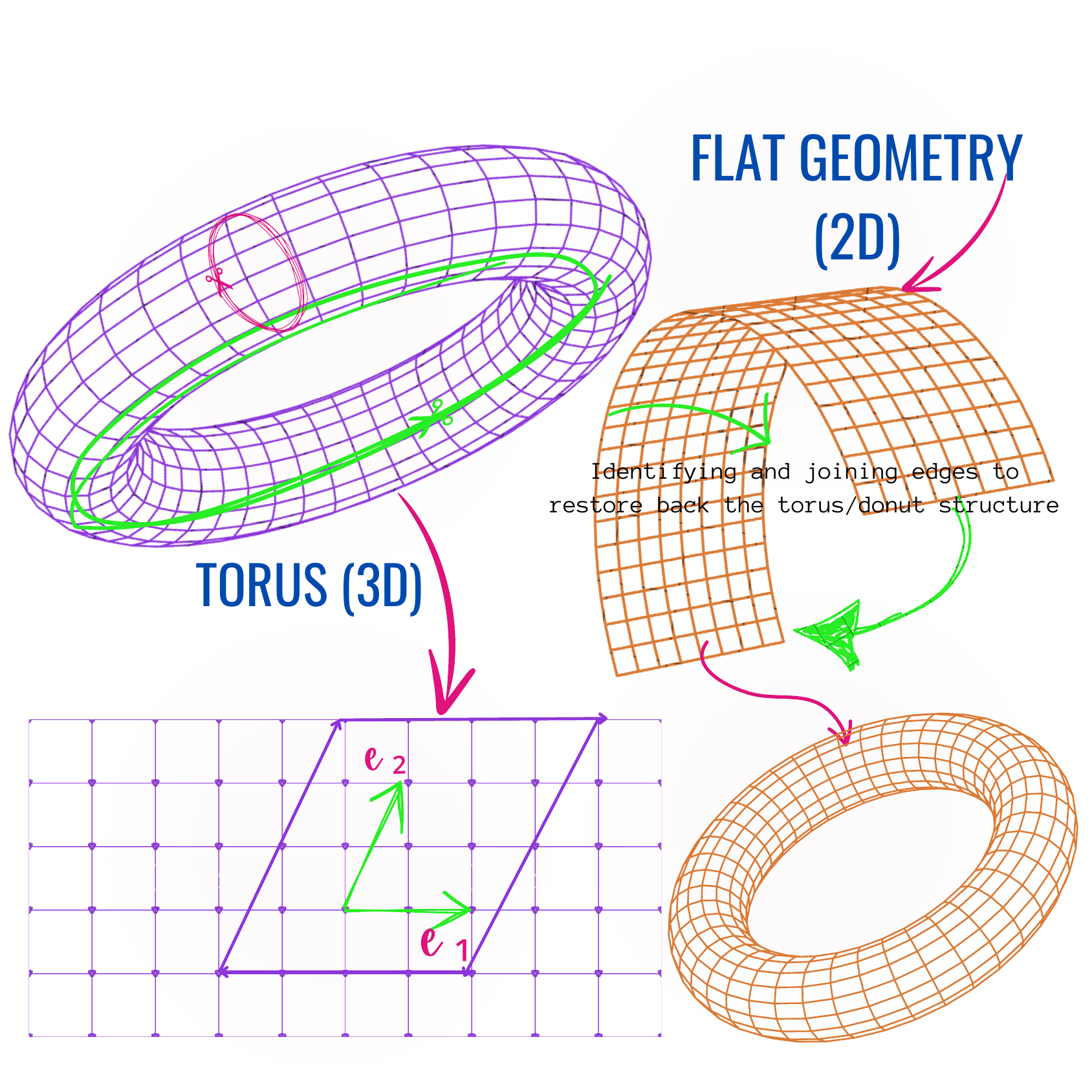}
\caption{Linear transformation of a Torus to flat geometry as a realisation of modular symmetry}
\label{fig1}
\end{figure}

It is well-know by now that the Standard Model, although being a very impressive theory in classifying particles in the pre-existing particle zoo, falls short in explaining the flavor structure paradigm of neutrino mixing and the origin of neutrino masses. As a consequence, theoretical physicists proposed a plethora of neutrino models based on traditional symmetries such as $A_{4}$, $A_{5}$, $S_{4}$, etc., to incorporate them as flavor symmetries in lepton mixing.\\
\\
However, when a non-Abelian finite discrete group like $A_{4}$ is imposed on the assigned Lagrangian of a neutrino model, the $A_{4}$ symmetry will eventually be broken down by Higgs-like scalar fields called Flavon fields, thus breaking the flavor symmetry. Following this spontaneous symmetry breaking, the resulting effective neutrino mass matrix will include three complexes or six real parameters such that the setup is invariant under the $A_{4}$ group.\\
\\
Nevertheless, this scenario is completely different when a modular symmetry is chosen to replace the flavor symmetry in model building. For instance, if a model is based on a finite $A_4$ modular group, then the Yukawa coupling is replaced by the modular forms. The Yukawa couplings, denoted as $ Y^{(k)}_{3} = \left( 
Y_{1},Y_{2}, Y_{3}\right)^{T} $ with weight k, transform as a triplet under the $ A_{4} $ modular group. There can be relations among $ Y_{1} $, $ Y_{2} $, $ Y_{3} $, but all of them will reduce to a single real parameter, namely the modulus parameter  $ \tau $, as shown later in Appendix \ref{app:A}. Consequently, the neutrino mass matrix will eventually have elements that are controlled by a single parameter  $ \tau $ instead of six real parameters.\\
\\
However, we have only discussed the class of modular symmetries with degree 1  \cite{Feruglio:2017spp}, \cite{Liu:2019khw} so far. This class of symmetries was first introduced by Feruglio Feruccio in his work \cite{Feruglio:2017spp} for neutrino model building. The modular symmetry arises from the Special Linear group, SL(2,$ \mathcal{Z} $), which is a  $ 2\times 2 $ matrix with integers as elements and unit determinant. The integers can be both positive or negative. The geometry of such an  SL(2,$ \mathcal{Z} $) modular symmetry can be visualized as the transformation of a torus in 3-dimension (3D). When this torus is cut along the two cycles as shown in Fig. \ref{fig1}, it will transform into a parallelogram in two dimensions (2D)  \cite{Ferrara:1989bc, Ferrara:1989qb, Kikuchi:2020frp, Almumin:2021fbk}.  Now, if we rejoin the edges of this flat geometry, it will restore the original shape of the torus. Thus, one can say that the symmetry related to the orbitfold or transformation of one set of basis vectors in a lattice to another in a compactified space can be realized exclusively by the moduli fields. \\
\\
In our work, we will be using modular forms of degree (or genus), $g>1$ to form a neutrino model.  We have specifically chosen a finite $ S_{4} \times Z_{2}$ Siegel modular group where the modular forms have degree, $g=2$. Similar to the $ SL(2, \mathcal{Z}) $ modular form, the $ Sp(2, \mathcal{Z})$ Siegel modular group is a symplectic group with a $ 2\times 2 $ matrix. However, the elements here are all matrices unlike $ SL(2, \mathcal{Z}) $ which consists of integer elements.  The Siegel modular group was first introduced in 1935 by Carl Ludwig Siegel \cite{Siegel}.  Later, Jun-Ichi Igusa extended the Siegel modular forms to higher degree for g=2 \cite{Igusa1962, Igusa1960, Igusa1964, Igusa1967, Igusa1979} and Shigeaki Tsuyumine did for g=3 \cite{Tsuyumine}. \\
\\
The Calabi-Yau manifold was motivated by the Siegel modular group \cite{Ishiguro:2021ccl, Ishiguro:2020nuf, Baur:2020yjl, Cecotti:1988qn, Cecotti:1988ad, Cecotti:1989kn, Dixon:1989fj, Candelas:1990pi, Strominger:1990pd, Ferrara:1991uz, Font:1992uk, Nilles:2021glx}which was one of the first applications of the Siegel modular group in physics. However, it finds its application in various subfields of physics and mathematics such as String Theory and Supersymmetry  \cite{Mayr:1995rx, Stieberger:1998yi}, Conformal Field Theory \cite{Cecotti:1988qn}, Number theory \cite{Ferrari:2018aiy, Geer:2023, Geer:2008, Bergstrom:2008, Siegel:1980, Zagier:1985, Anatoli:2008, Murty:2016, Koblitz:1993, Ono:2004, Meher:2021, Meher:2012, Ramakrishnan, Abhas:2020, Kohnen:1991}, etc.  For interested readers who want to have an introductory knowledge of the mathematics of Siegel, one can follow the literature given in \cite{Ding:2020zxw, Ding:2021iqp, Tatsuishi:2019osz, Ding:2023htn,Kikuchi:2023dow}. Some models based on $ A_{4} $, $ S_{4} $, eclectic, and other flavour groups can be found in \cite{Chen:2024otk, Nilles:2023shk, Kobayashi:2023zzc, deMedeirosVarzielas:2023crv, Meloni:2023aru, Ding:2019zxk, Novichkov:2018nkm, Ding:2019xna, Liu:2020akv, Novichkov:2020eep, Liu:2020msy, Yao:2020zml, Wang:2020lxk, Ding:2020msi, Li:2021buv, Ding:2023ydy,  Arriaga-Osante:2023wnu, Petcov:2023fwh, Chen:2019ewa, Chen:2021prl, CentellesChulia:2023osj, Nilles:2020kgo, Nilles:2020nnc, Ding:2023ynd, Kumar:2023moh, Baur:2020jwc, Kashav:2022kpk, Kashav:2021zir, Nomura:2023usj, Mishra:2023ekx, deAnda:2023udh, Mishra:2023cjc, Devi:2023vpe}. Few works on some interesting symmetries such as magic symmetry, $\mu$-$\tau$ reflection symmetry are also currently ongoing, which can be found in \cite{ Singh:2022nmk, Singh:2022tvz, Chakraborty:2023msb, Zhou:2014sya} \\
\\
To summarize, our paper has been organized as follows. In section \ref{sec1}, we provide a  brief overview on the Siegel Modular group and Siegel Modular forms. In section\ref{sec2}, we introduce our neutrino model based on $ S_{4}\times Z_{2} $ finite modular group, and in section \ref{sec3}, its results obtained from our model are presented and discussed. In the end, we offer a gist of our findings in our conclusion section \ref{Sec4}. 
\subsection{Siegel Modular Group and Siegel Modular Forms}
\label{sec1}

The Siegel modular forms were introduced by C. L. Siegel in the year 1935 in \enquote{Uber die analytische Theorie der quadratischen Formen}, Ann. Math \cite{Siegel}. Siegel modular forms are the generalization of the classical modular forms in  SL(2, $\mathcal{Z}$).  The upper half plane of the classical modular forms with g=1 is now replaced by the Siegel upper half plane  $\mathcal{H}_g$, and the automorphism group Sp(2g, $\mathcal{Z}$)  of a unimodular symplectic form on $\mathcal{Z}^{2g}$ replaces the usual SL(2, $\mathcal{Z}$) group. The integer $g$, which is usually called genus or degree, has a value of $g>1$ for the Siegel modular forms \cite{Geer:2008}.Unlike the classical modular group, which has the modulus  $ \tau $ as the only free parameter, the Siegel modular group has more than one variable. The space of the Siegel upper half plane can be denoted as \cite{Zagier:1985}:\\
\begin{equation}
\mathcal{H}_{g}=\left\lbrace  \mathcal{\tau} \epsilon \; Mat(g \times g, \mathcal{C})| \mathcal{\tau}=\tau^{t} \textrm{ } \& \textrm{ } ,  Im(\mathcal{\tau})>0\right\rbrace 
\end{equation} 
where $ g\times g $ represents complex symmetric matrices with positive definite imaginary part of the every matrix entry. 
\\
The  symplectic group Sp(2, $ \mathcal{Z} $)  is often denoted as $ \Gamma_{g} $ which is called Siegel modular group of degree g. This group satisfies the symplectic condition defined by \cite{Zagier:1985}\\
\begin{equation}
Sp_{2n}(\mathbb{R}) = \left\lbrace  M \epsilon M_{2n}(\mathbb{R}) \mid M J_{2n}M^{t}=J_{2n}\right\rbrace \textrm{ , }  J_{2n}= \left(\begin{matrix}
0_{n} & -I_{n}\\
-I_{n} & 0_{n}
\end{matrix}\right)
\end{equation}
where, $ 0_{n}=[a_{ij}=0]_{n\times n} $ and $ I_{n}=\begin{bmatrix}a_{ij}=1 , \textrm{ if } i=j \\ a_{ij}=0 , \textrm{ if } i\neq j \end{bmatrix}_{n\times n} $.
\\
Thus, we have
\begin{equation}
Sp_{2n}(\mathbb{R}) = \left\lbrace \left( \begin{matrix}
A & B\\
C & D
\end{matrix} \right) \mid A, B, C, D \;\; \epsilon \;\; M_{n}(\mathbb{R}),\;\; AB^{t}=BA^{t}, \; CD^{t}=DC^{t} \; \& \; AD^{t}- BC^{t}=I_{n} \right\rbrace 
\end{equation}
Thus for degree $ g\geq 2 $, a Siegel modular form F of weight \enquote{k} and level n is a function $ F  :\mathcal{H}_{g}\rightarrow \mathbb{C} $ satisfying the following conditions:\\\\
1. F is a holomorphic function on $ \mathcal{H}_{g} $\\
2. F(M$ \cdot  \mathcal{Z}$) = det $ (C\mathcal{Z} + D)^{k}F(\mathcal{Z}) $, 
for all $ M = \left( \begin{matrix}
 A & B\\
 C & D
\end{matrix} \right) \epsilon \; \Gamma_{n}, \; \mathcal{Z} \epsilon \mathcal{H}_{g} $. The $ \mathcal{H}_{g} $ transforms under $\Gamma_g(n)$ as \cite{Ding:2020zxw}
\begin{equation}
\label{SiegelForms}
f(\gamma \tau)=\det(C\tau+D)^k f(\tau) \,, \quad\quad \gamma=\begin{pmatrix}
A & B \\
C & D
\end{pmatrix} \in \Gamma_g(n) \,.
\end{equation}
\\
This Siegel modular group $ \Gamma_{g}$ has two generators $ S $ and $ T_{i} $ given as  \cite{Ding:2023htn}
\begin{equation}
S= \left( \begin{matrix}
0 & \mathbb{1}_{g}\\
-\mathbb{1}_{g} & 0
\end{matrix} \right), \; T_{i}=\left( \begin{matrix}
\mathbb{1}_{g} & B_{i}\\
0 & \mathbb{1}_{g}
\end{matrix} \right)
\end{equation} 
where $ B_{i} $ represents the basis for a $ g\times g $ symmetric mstrices. The generator S satisfies the symplectic relation, $ S^{2}= -\mathbb{1}_{2g}$. Without the loss of generality, we take B as chosen in the \cite{Ding:2023htn} for degree 2.
\begin{equation}
B_{1}= \left( \begin{matrix}
1 & 0\\
0 & 0
\end{matrix}\right) ,  \; B_{2}= \left( \begin{matrix}
0 & 0\\
0 & 1
\end{matrix}\right), \; B_{3}= \left( \begin{matrix}
0 & 1\\
1 & 0
\end{matrix}\right)
\end{equation}
Thus under the modularity condition of S and T, the modulus parameter $ \tau $ transforms as 
\begin{equation}
\tau \xrightarrow[]{\text{S}} -\tau^{-1}, \; \; \tau \xrightarrow[]{\text{$ T_{i} $}} \tau + B_{i}
\end{equation}
For degree g=1, the generators $ S $ and $ T_{i} $ will reduce back to the usual generators of a modular group on $ SL (2, \mathcal{Z}) $ space.

\subsection{An $S_4\times Z_2$ Siegel modular invariant model }
\label{sec2}
We develop an $S_4\times Z_2$ Siegel modular invariant model where we generate the neutrino masses from the type-I seesaw mechanism. We denote the left-handed lepton as $L$, right-handed neutrinos as $N^c$ and right-handed charged leptons as $E^c$. Under the $S_4\times Z_2$ Siegel  modular group, $L$ and $E^c$ transform as a triplet $\mathbf{3'}$ and the right-handed neutrinos $N^c$ transform as a triplet $\mathbf{3}$. The weights assigned to the fields are given below:
\begin{align}
\nonumber&\rho_{E^c}= \rho_{L} = \mathbf{3'},~~~~\rho_{N^c}=\mathbf{3},~~~~ \rho_{H_u}=\rho_{H_d}=\mathbf{1}\,,\\
&k_{H_u}=k_{H_d}=0,~~~ k_{E^c}=k_{N^c}=0,~ ~k_{L}=-2\,.
\end{align}
Thus, the superpotential for the charged lepton sector and the neutrino sector can be written as
\begin{align}
\nonumber w_e &=  \alpha (E^c L Y_\mathbf{3'})_\mathbf{1}H_d + \beta (E^c L Y_\mathbf{1})_\mathbf{1}H_d\,, \\
 w_\nu &= g_1 (N^c L Y_\mathbf{3'})_\mathbf{1} H_u   +  \Lambda(N^c N^c)_\mathbf{1}\,.
 \label{Lag}
\end{align}
Using the Clebsch-Gordon coefficients of the $S_4\times Z_2$ symmetry group listed in Appendix~\ref{app:A}, we get the charged lepton and neutrino mass matrices:
\begin{equation}
\begin{aligned}
\label{lept1}
M_e=  \begin{pmatrix}
2 \alpha Y_1 + \beta Y_4 & -\alpha Y_3 & -\alpha Y_2 \\
-\alpha Y_3 & 2\alpha Y_2 & -\alpha Y_1+ \beta Y_4 \\
-\alpha Y_2 & -\alpha Y_1 + \beta Y_4 & 2\alpha Y_3
\end{pmatrix}v_d
\end{aligned}
\end{equation}
\begin{equation}
\begin{aligned}
M_D=\left(
\begin{array}{ccc}
 0 & -g_1 v_u Y_3 & g_1 v_u Y_2 \\
 g_1 v_u Y_3 & 0 & -g_1 v_u Y_1 \\
 -g_1 v_u Y_2 & g_1 v_u Y_1 & 0 \\
\end{array}
\right)\end{aligned}
\end{equation}
\begin{equation}
\begin{aligned}
M_N= \left(
\begin{array}{ccc}
 \Lambda  & 0 & 0 \\
 0 & 0 & \Lambda  \\
 0 & \Lambda  & 0 \\
\end{array}
\right)\,.
\end{aligned}
\end{equation}
The light neutrino mass matrix $m_{\nu}$ is given by the seesaw formula
\begin{equation}
M_\nu= -M_D^T M_N^{-1} M_D\,.
\end{equation}
\begin{equation}
M_\nu=\left(
\begin{array}{ccc}
 2 Y_2 Y_3 & -Y_1 Y_3 & -Y_1 Y_2 \\
 -Y_1 Y_3 & -Y_3^2 & Y_1^2+Y_2 Y_3 \\
 -Y_1 Y_2 & Y_1^2+Y_2 Y_3 & -Y_2^2 \\
\end{array}
\right)\frac{g_1^2 v_u^2}{\Lambda }
\end{equation}

It is interesting to note that our model does not demand any additional flavon or any residual symmetry for the model building purpose. Also, the model does not require modular forms with higher weights. However, one can always explore  the analysis of a model with higher weights, incorporating additional symmetries and flavons to find its implications in neutrino oscillation. \\
\\
For the input parameters, i.e., the CPV phase $ \delta $, the mixing angles and the mass-squared differences, we have utilized the latest NuFIT 5.2 \cite{Esteban:2020cvm} database.

\section{Results and Discussion}
\label{sec3}

\begin{figure}[t!]
\centering
\includegraphics[width=0.7\linewidth]{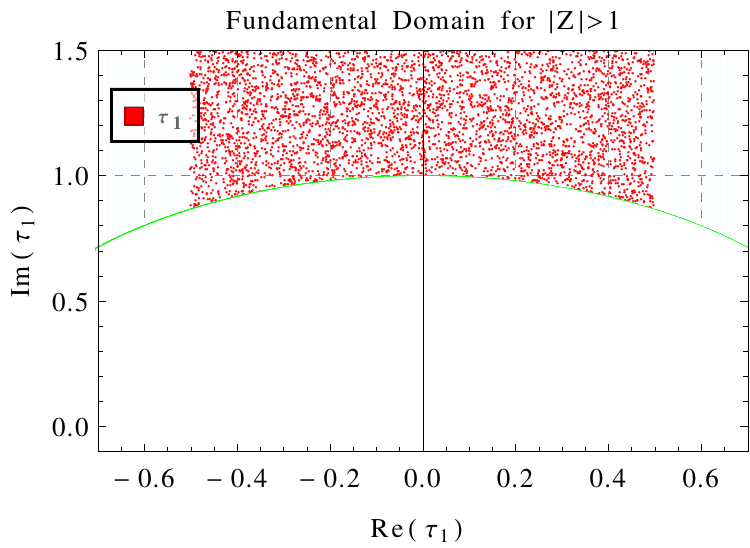}
\caption{ Correlation between $Re (\tau_{1} )$ and $Im(\tau_{1} )$ within the fundamental domain}
\label{fig2}
\end{figure}

In our model we have computed the modulus parameters $\tau_1$ and $\tau_3$, as well as other neutrino oscillation parameters. The optimal values of $\tau_1$ and $\tau_3$ obtained from our model are,

\begin{equation}
\dfrac{v^{2}_{u}g^{2}_{1}}{\Lambda}=246\times 10^{-9}\; eV, \;  \; \tau_{1}= -0.00138716+1.30102 i , \; \; \tau_{3}=-0.00337176+1.30109 i
\label{opt}
\end{equation}

\begin{figure}[t!]
\centering
\includegraphics[width=0.7\linewidth]{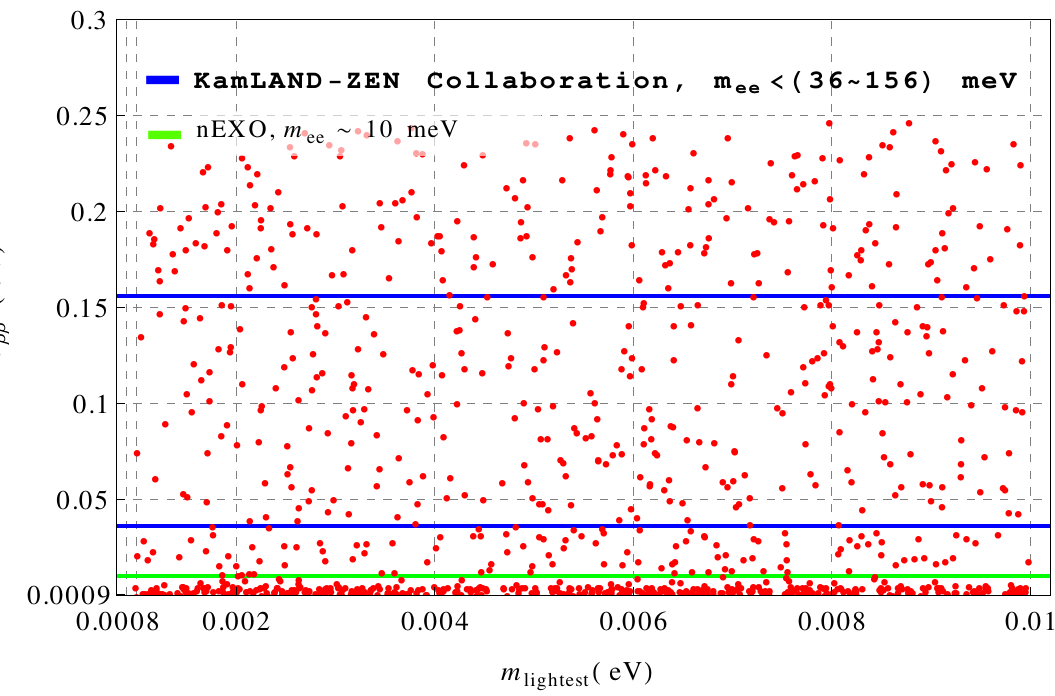}
\caption{ Correlation between $m_{lightest}$ and $m_{\beta \beta}$}
\label{fig3}
\end{figure}

\begin{figure}[t!]
\centering
\includegraphics[width=0.62\linewidth]{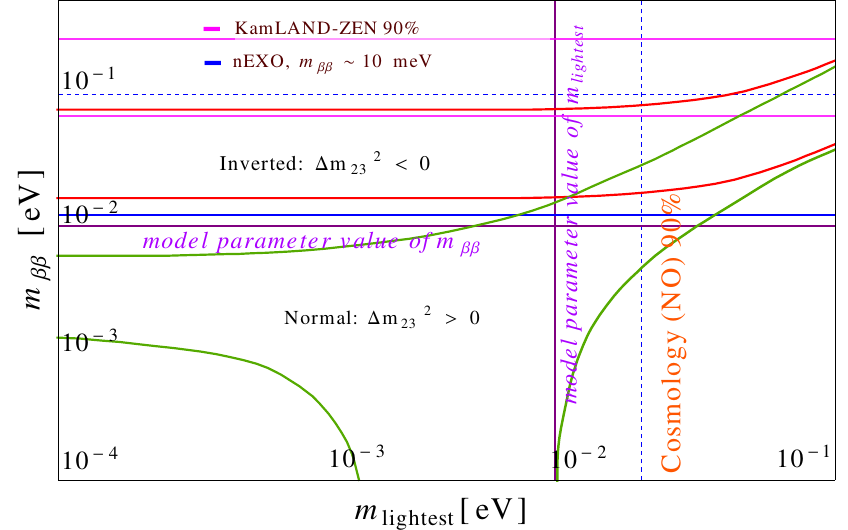}
\caption{ In this correlation plot of $m_{lightest}$ - $m_{\beta \beta}$, we highlight the limit of the parameters  $m_{lightest}$ and $m_{\beta \beta}$ as obtained from our model and compare them with the stringest limits given by KamLAND-ZEN Collaboration and nEXO experiments. }
\label{fig5}
\end{figure}

\begin{figure}[t!]
\centering
\includegraphics[width=0.7\linewidth]{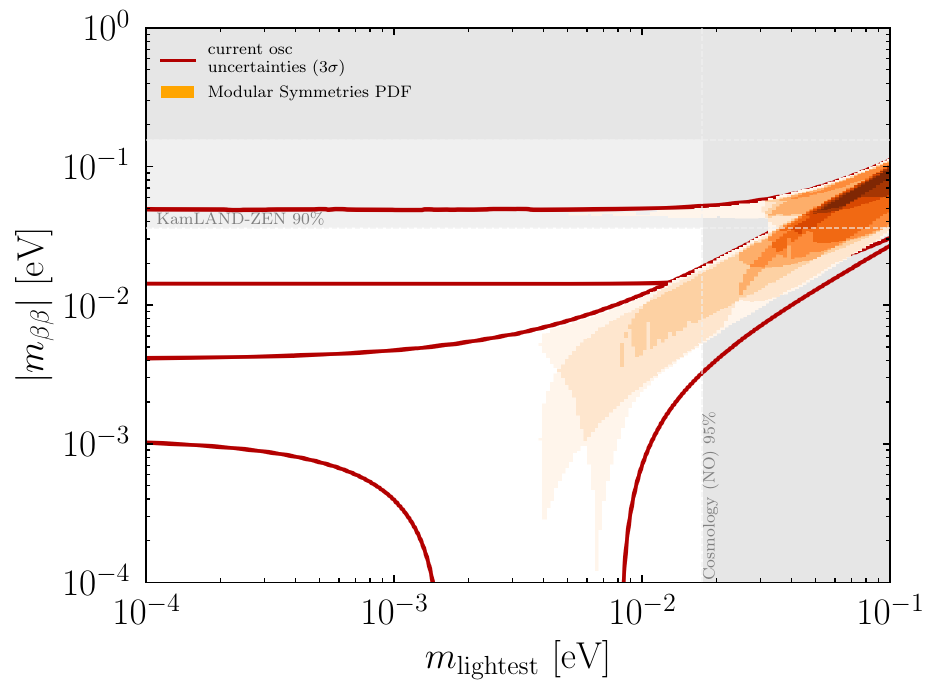}
\caption{ The Probability Density Function [PDF] of models with modular symmetries using the latest constraints from the oscillation data \cite{Denton:2023hkx}}
\label{fig6}
\end{figure}

\begin{figure}[t!]
\centering
\includegraphics[width=0.7\linewidth]{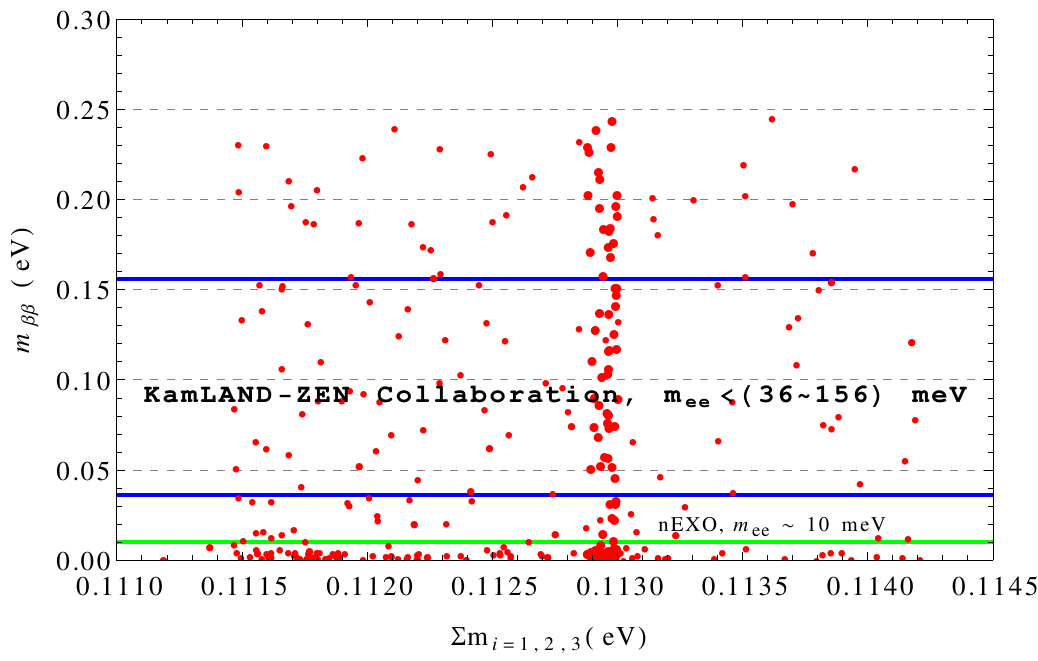}
\caption{ Correlation between $\Sigma m_{i=1,2,3}$ and $m_{\beta \beta}$}
\label{fig4}
\end{figure}

In Figure \ref{fig2}, we have presented the variation of $Re(\tau_1)$ versus $Im(\tau_1)$ within the fundamental domain, which shows the allowed region of $ \tau_{1} $ in the fundamental domain of $ Sp(2,\mathcal{Z}) $. From our model, we have predicted the values of linearly independent Yukawa couplings with weight k=2, as well as the effective neutrino mass of the neutrinoless double beta decay, $ m_{\beta \beta} $. Figure \ref{fig3} demonstrates that for optimal values of the Yukawa coupling  constants $ Y_{2} $ and $ Y_{3} $, the calculated values of  $ m_{\beta \beta} $  align with the previously given limit $ m_{\beta \beta}<(61 - 165) $  meV as provided by KamLAND-ZEN experiment in \cite{KamLAND-Zen:2016pfg} and its latest limit $ m_{\beta \beta}<(36 - 156) $ \cite{KamLAND-Zen:2022tow, Denton:2023hkx}. Also, for certain points as shown in the same plot, the model value of $ m_{\beta \beta}$ aligns with future sensitivity value of $ m_{\beta \beta}\sim \; 10\; meV$, as given by nEXO experiment \cite{nEXO:2017nam}. In ref. \cite{Denton:2023hkx}, the authors studied different flavor models, including some modular forms with degree, g=1 which is based on \cite{Gehrlein:2020jnr}. Specifically they looked at which models lead to very low predicted $ \nu \beta \beta $ rates in the funnel. It is interesting to note that that the allowed region for the Probability Density Function (PDF)  of  $\vert m_{\beta \beta} \vert$ - $ m_{lightest} $ as shown in Figure \ref{fig6} in ref. \cite{Denton:2023hkx}, agrees well with the allowed region of $\vert m_{\beta \beta}\vert$ - $ m_{lightest} $ plot in Figure \ref{fig5} as obtained from our model with degree, g=2, with substantial favour for normal ordering in both cases.  \\
\\
In work \cite{Vagnozzi:2017ovm}, the authors together with the team from the Planck and BOSS collaborations, provided state-of-the-art constraints on neutrino properties using a wide range of data. They were the pioneers to obtain the $ \Sigma m_{i} < 0.12 \; eV$  upper limit  \cite{Vagnozzi:2017ovm} and it was later the  upper bound was obtained from CMB data by Planck after 1.5 years \cite{Planck:2018vyg}. In the same work, they authors reported some of the first hints for a preference for the normal ordering, with similar results of the BAO data as extracted from the MGS, 6dFGS, and BOSS DR12 galaxy surveys \cite{Tanseri:2022zfe}. Moreover, later in \cite{Vagnozzi:2018jhn}, it was shown on how this bound can get tighter in well-motivated models of dark energy, contrary to the naive expectations given that the parameter space is enlarged. Cosmology is in principle also able to constrain the mass ordering, and recent cosmological observations have been argued to slightly favor the normal mass ordering \cite{RoyChoudhury:2019hls, Hergt:2021qlh, Jimenez:2022dkn}.\\
 \\
From Figure \ref{fig4}, we can conclude that the sum of the light neutrino masses satisfies the cosmological constrain which is $ \Sigma m_{i} < 0.12 \; eV$ \cite{Planck:2018vyg}. Interestingly,  This confirms the viability of our model and accuracy of the model parameters values.\\
\\
To obtain accurate and predictable values of the unknown neutrino oscillation parameters, such as $ \delta $, $ m_{lightest} $, we have utilized the data of the mixing angles and mass-squared differences as input parameters from the NuFIT 5.2 \cite{Esteban:2020cvm} database within the 3$ \sigma $ range. It is worth mentioning that our model does not incorporate any flavon so far and it is quite feasible for testability in the neutrino experiments. Furthermore, the findings of our model offer valuable insight in the search of neutrinoless double beta decay in future and ongoing experiments such as KamLAND-ZEN collaboration, nEXO etc.

\section{Conclusion}
\label{Sec4}
In our work, we use an $ S_{4} \times Z_{2}$ finite Siegel modular form. We constrain the modulus space of the matrix $ \tau $ for $\tau_{1}=\tau_{2}  $. This gives us four linearly independent modular forms with weight 6, instead of five. With this framework, we design a neutrino model based on Type-I seesaw mechanism to generate the light neutrino masses. By optimizing the values of $ \tau_{1} $, $ \tau_{2} $ and  $ \dfrac{v^{2}_{u}g^{2}_{1}}{\Lambda} $ as given in Equation\ref{opt}, the values of the Yukawa coupling constants with weight 2 are determined as given in the Lagrangian [Equation\ref{Lag}]. These data are further utilized to determine the values of the lightest neutrino mass, neutrino mass for neutrinoless double beta decay and the sum of the light neutrino masses. For the values of the input parameters like the mass-squared differences and the mixing angles, we take the data from the latest NuFIT 5.2 \cite{Esteban:2020cvm}\\
\\  
It is interesting to find that our model can successfully predict the effective neutrino mass of  neutrinoless double beta decay within the stringent limit of KamLAND-ZEN Collaboration \cite{KamLAND-Zen:2016pfg}. Also some values of $ m_{\beta \beta} $ aligns well with its experimental value provided by nEXO experiment \cite{nEXO:2017nam}. To ensure the accuracy of our measured parameters, we find that the sum of the light neutrino masses from our model agrees well with the cosmological constrain. Thus, our model exhibits high testability in the future collider experiments.
\\
\\
\\
\newpage
\textbf{Acknowledgements}\\
\\
The author expresses sincere gratitude to  Feruglio Ferrucio (INFN, Sezione di Padova, Italia), Monal Kashav (Department of Physics and Astronomical Science, Central University of Himachal Pradesh), Labh Singh (Department of Physics and Astronomical Science, Central University of Himachal Pradesh), Ranjeet Kumar (IISER, Bhopal) and Peter B. Denton (Brookhaven National Laboratory, USA)  for their valuable feedback and discussions. Overall, the author is deeply appreciative of the support and knowledge provided by these esteemed individuals, which has undoubtedly contributed to the success of this research work.
\begin{appendices}
\setlength{\extrarowheight}{0.0 cm}


\section{The finite Siegel modular group $S_4\times Z_2$}
\label{app:A}
The finite modular group $S_4\times Z_2$ has three generators that satisfies the following relation \cite{Ding:2020zxw} :
\begin{equation}
\mathcal{S}^2=\mathcal{T}^3=(\mathcal{S}\mathcal{T})^4=1,~~\mathcal{V}^2=1,~~~\mathcal{S}\mathcal{V}=\mathcal{V}\mathcal{S},~~~
\mathcal{T}\mathcal{V}=\mathcal{V}\mathcal{T}\,.
\end{equation}
There are four singlet representations $\mathbf{1}$, $\mathbf{1}'$, $\mathbf{\hat{1}}$, $\mathbf{\hat{1}'}$, two doublet representations $\mathbf{2}$, $\mathbf{\hat{2}}$, and four triplet representations $\mathbf{3}$, $\mathbf{3'}$, $\mathbf{\hat{3}}$ and $\mathbf{\hat{3}'}$ in the $S_4\times Z_2$.  The multiplication rules of these irreducible representations in $S_4\times Z_2$ are given as \cite{Ding:2020zxw},
\begin{align}
\nonumber&\mathbf{1'} \otimes \mathbf{1'} =\mathbf{\hat{1}} \otimes \mathbf{\hat{1}} =\mathbf{\hat{1}'} \otimes \mathbf{\hat{1}'} =\mathbf{1},~\mathbf{1'} \otimes \mathbf{\hat{1}} =\mathbf{\hat{1}'},~\mathbf{1'} \otimes \mathbf{\hat{1}'} =\mathbf{\hat{1}},~\mathbf{\hat{1}} \otimes \mathbf{\hat{1}'} =\mathbf{1'}\,,\\
\nonumber&\mathbf{1'} \otimes \mathbf{2} =\mathbf{\hat{1}'} \otimes \mathbf{\hat{2}}=\mathbf{\hat{1}} \otimes \mathbf{\hat{2}}=\mathbf{2},~
\mathbf{1'} \otimes \mathbf{\hat{2}} =\mathbf{\hat{1}'} \otimes \mathbf{2}=\mathbf{\hat{1}} \otimes \mathbf{2}=\mathbf{\hat{2}}\,,\\
\nonumber&\mathbf{1'} \otimes \mathbf{3'} = \mathbf{\hat{1}} \otimes \mathbf{\hat{3}}=\mathbf{\hat{1}'} \otimes \mathbf{\hat{3}'}=\mathbf{3}, ~\mathbf{1'} \otimes \mathbf{3} = \mathbf{\hat{1}} \otimes \mathbf{\hat{3}'}=\mathbf{\hat{1}'} \otimes \mathbf{\hat{3}}=\mathbf{3'}, \\
\nonumber&\mathbf{1'} \otimes \mathbf{\hat{3'}} = \mathbf{\hat{1}} \otimes \mathbf{3}=\mathbf{\hat{1}'} \otimes \mathbf{3'}=\mathbf{\hat{3}}, ~\mathbf{1'} \otimes \mathbf{\hat{3}} = \mathbf{\hat{1}} \otimes \mathbf{3'}=\mathbf{\hat{1}'} \otimes \mathbf{3}=\mathbf{\hat{3}'}\,,\\
\nonumber&\mathbf{2} \otimes \mathbf{2} =\mathbf{\hat{2}} \otimes \mathbf{\hat{2}} =\mathbf{1}\oplus\mathbf{1'}\oplus\mathbf{2},~\mathbf{2} \otimes \mathbf{\hat{2}} =\mathbf{\hat{1}}\oplus\mathbf{\hat{1}'}\oplus\mathbf{\hat{2}}\,,\\
\nonumber&\mathbf{2} \otimes \mathbf{3} =\mathbf{2} \otimes \mathbf{3'} =\mathbf{3}\oplus\mathbf{3'},~\mathbf{2} \otimes \mathbf{\hat{3}} =\mathbf{2} \otimes \mathbf{\hat{3}'} =\mathbf{\hat{3}}\oplus\mathbf{\hat{3}'},\\
\nonumber&\mathbf{\hat{2}} \otimes \mathbf{3}=\mathbf{\hat{2}} \otimes \mathbf{3'}=\mathbf{\hat{3}}\oplus\mathbf{\hat{3}'},~\mathbf{\hat{2}} \otimes \mathbf{\hat{3}}=\mathbf{\hat{2}} \otimes \mathbf{\hat{3}'}  =\mathbf{3}\oplus\mathbf{3'}\,,\\
\nonumber&\mathbf{3} \otimes \mathbf{3}=\mathbf{3'} \otimes \mathbf{3'}=\mathbf{\hat{3}} \otimes \mathbf{\hat{3}}=\mathbf{\hat{3}'} \otimes \mathbf{\hat{3}'}=\mathbf{1}\oplus\mathbf{2}\oplus\mathbf{3}\oplus\mathbf{3'}, \\
\nonumber&\mathbf{3} \otimes \mathbf{3'}=\mathbf{\hat{3}} \otimes \mathbf{\hat{3}'}=\mathbf{1'}\oplus\mathbf{2}\oplus\mathbf{3}\oplus\mathbf{3'}, \\
\nonumber&\mathbf{3} \otimes \mathbf{\hat{3}}=\mathbf{3'} \otimes \mathbf{\hat{3}'}=\mathbf{\hat{1}}\oplus\mathbf{\hat{2}}\oplus\mathbf{\hat{3}}\oplus\mathbf{\hat{3}'}, \\
&\mathbf{3} \otimes \mathbf{\hat{3}'}=\mathbf{3'} \otimes \mathbf{\hat{3}}=\mathbf{\hat{1}'}\oplus\mathbf{\hat{2}}\oplus\mathbf{\hat{3}}\oplus\mathbf{\hat{3}'}\,.
\end{align}

The Clebsch-Gordan coefficients for product of two representations with correspoding elements $\alpha_i$ and  $\beta_i$ are shown in Table~\ref{tab:1} \cite{Ding:2020zxw}.

\begin{table}[ht!]
\centering
\resizebox{1.0\textwidth}{!}{
\begin{tabular}{|c|c|c|c|c|c|c|c|c|c|c|c|c|c|}\hline\hline
\multicolumn{3}{|c}{~~~$\mathbf{1} \otimes \mathbf{2}= \mathbf{\hat{1}} \otimes \mathbf{\hat{2}} = \mathbf{2}~~~$} & \multicolumn{3}{|c}{~$\mathbf{1} \otimes \mathbf{\hat{2}}= \mathbf{\hat{1}} \otimes \mathbf{2} = \mathbf{\hat{2}}~$} & \multicolumn{3}{|c}{~~~$\mathbf{1'} \otimes \mathbf{2}= \mathbf{\hat{1}'} \otimes \mathbf{\hat{2}} = \mathbf{2}~~~$}&\multicolumn{3}{|c|}{$~~\mathbf{1'} \otimes \mathbf{\hat{2}} = \mathbf{\hat{1}'} \otimes \mathbf{2} = \mathbf{\hat{2}}~~$} \rule[-0.5ex]{-4pt}{3ex} \\ \hline
\multicolumn{3}{|c}{  $\mathbf{2}\sim\begin{pmatrix}
 \alpha\beta_1 \\
 \alpha\beta_2 \\
\end{pmatrix} $} &
\multicolumn{3}{|c}{$\mathbf{\hat{2}}\sim \begin{pmatrix}
 \alpha\beta_1 \\
 \alpha\beta_2 \\
\end{pmatrix} $} &
\multicolumn{3}{|c}{$\mathbf{2}\sim\begin{pmatrix}
 -\alpha\beta_1 \\
 \alpha\beta_2 \\
\end{pmatrix} $} &
\multicolumn{3}{|c|}{$\mathbf{\hat{2}}\sim\begin{pmatrix}
 -\alpha\beta_1 \\
 \alpha\beta_2 \\
\end{pmatrix} $} \rule[-3ex]{-4pt}{7ex} \\ \hline\hline

\multicolumn{6}{|c}{~~$\mathbf{1} \otimes \mathbf{3} = \mathbf{1'} \otimes \mathbf{3'} = \mathbf{\hat{1}} \otimes \mathbf{\hat{3}} = \mathbf{\hat{1}'} \otimes \mathbf{\hat{3}'}=\mathbf{3}~~$} & \multicolumn{6}{|c|}{~~$\mathbf{1} \otimes \mathbf{3'} = \mathbf{1'} \otimes \mathbf{3} = \mathbf{\hat{1}} \otimes \mathbf{\hat{3}'} = \mathbf{\hat{1}'} \otimes \mathbf{\hat{3}}=\mathbf{3'}~~$} \rule[-0.5ex]{-4pt}{3ex}  \\ \hline
\multicolumn{6}{|c}{$\mathbf{3}\sim\begin{pmatrix}
 \alpha\beta_1 \\
 \alpha\beta_2 \\
 \alpha\beta_3 \\
\end{pmatrix} $}&
\multicolumn{6}{|c|}{$ \mathbf{3'}\sim\begin{pmatrix}
 \alpha\beta_1 \\
 \alpha\beta_2 \\
 \alpha\beta_3 \\
\end{pmatrix} $} \rule[-4.5ex]{0pt}{10ex}\\ \hline
  \multicolumn{6}{|c}{~~$\mathbf{1} \otimes \mathbf{\hat{3}} = \mathbf{1'} \otimes \mathbf{\hat{3}'} = \mathbf{\hat{1}} \otimes \mathbf{3} = \mathbf{\hat{1}'} \otimes \mathbf{3'} = \mathbf{\hat{3}}~~$} & \multicolumn{6}{|c|}{~~$\mathbf{1} \otimes \mathbf{\hat{3}'} = \mathbf{1'} \otimes \mathbf{\hat{3}} = \mathbf{\hat{1}} \otimes \mathbf{3'} = \mathbf{\hat{1}'} \otimes \mathbf{3} = \mathbf{\hat{3}'}~~$} \rule[-0.5ex]{-4pt}{3ex} \\ \hline
\multicolumn{6}{|c}{$\mathbf{\hat{3}}\sim\begin{pmatrix}
 \alpha\beta_1 \\
 \alpha\beta_2 \\
 \alpha\beta_3 \\
\end{pmatrix}$ } &
\multicolumn{6}{|c|}{$\mathbf{\hat{3}'}\sim\begin{pmatrix}
 \alpha\beta_1 \\
\alpha\beta_2 \\
 \alpha\beta_3 \\
\end{pmatrix}$ } \rule[-4.5ex]{-4pt}{10ex} \\ \hline\hline

\multicolumn{4}{|c}{~~~~~~~~~$\mathbf{2} \otimes \mathbf{2} = \mathbf{1_s} \oplus \mathbf{1'_a} \oplus \mathbf{2_s}$~~~~~~~~~} & \multicolumn{4}{|c}{~~~~~~~~~~$\mathbf{2} \otimes \mathbf{\hat{2}} = \mathbf{\hat{1}} \oplus \mathbf{\hat{1}'} \oplus \mathbf{\hat{2}}$~~~~~~~~~~~} & \multicolumn{4}{|c|}{$\mathbf{\hat{2}} \otimes \mathbf{\hat{2}} = \mathbf{1_s} \oplus \mathbf{1'_a} \oplus \mathbf{2_s}$ } \rule[-0.5ex]{-4pt}{3ex}  \\ \hline
 \multicolumn{4}{|c}{  $ \begin{array}{l}
 \mathbf{1_s}\sim \alpha_1 \beta_2+\alpha_2 \beta_1 \\
 \mathbf{1'_a}\sim \alpha_1 \beta_2-\alpha_2 \beta_1 \\
 \mathbf{2_s}\sim\begin{pmatrix}
 \alpha_2 \beta_2 \\
 \alpha_1 \beta_1 \\
\end{pmatrix} \\
\end{array} $ } &
\multicolumn{4}{|c}{  $\begin{array}{l}
 \mathbf{\hat{1}}\sim \alpha_1 \beta_2+\alpha_2 \beta_1 \\
 \mathbf{\hat{1}'}\sim \alpha_1 \beta_2-\alpha_2 \beta_1 \\
 \mathbf{\hat{2}}\sim\begin{pmatrix}
 \alpha_2 \beta_2 \\
 \alpha_1 \beta_1 \\
\end{pmatrix} \\
\end{array} $} &
\multicolumn{4}{|c|}{ $ \begin{array}{l}
 \mathbf{1_s}\sim \alpha_1 \beta_2+\alpha_2 \beta_1 \\
 \mathbf{1'_a}\sim \alpha_1 \beta_2-\alpha_2 \beta_1 \\
 \mathbf{2_s}\sim\begin{pmatrix}
 \alpha_2 \beta_2 \\
 \alpha_1 \beta_1 \\
\end{pmatrix} \\
\end{array} $} \rule[-6ex]{-4pt}{13ex} \\ \hline\hline

\multicolumn{3}{|c}{~$\mathbf{2} \otimes \mathbf{3} = \mathbf{\hat{2}} \otimes \mathbf{\hat{3}} = \mathbf{3} \oplus \mathbf{3'} $} & \multicolumn{3}{|c}{$\mathbf{2} \otimes \mathbf{3'} = \mathbf{\hat{2}} \otimes \mathbf{\hat{3}'} =\mathbf{3} \oplus \mathbf{3'}~$} & \multicolumn{3}{|c}{~$\mathbf{2} \otimes \mathbf{\hat{3}} = \mathbf{\hat{2}} \otimes \mathbf{3} = \mathbf{\hat{3}} \oplus \mathbf{\hat{3}'}~$}& \multicolumn{3}{|c|}{~$\mathbf{2} \otimes \mathbf{\hat{3}'}= \mathbf{\hat{2}} \otimes \mathbf{3'} = \mathbf{\hat{3}} \oplus \mathbf{\hat{3}'}~$} \rule[-0.5ex]{-4pt}{3ex} \\ \hline
 \multicolumn{3}{|c}{ $\begin{array}{l}
 \mathbf{3}\sim\begin{pmatrix}
 \alpha_2 \beta_3 + \alpha_1 \beta_2  \\
 \alpha_2 \beta_1 + \alpha_1 \beta_3  \\
 \alpha_2 \beta_2 + \alpha_1 \beta_1  \\
\end{pmatrix}  \\
 \mathbf{3'}\sim\begin{pmatrix}
 \alpha_2 \beta_3 - \alpha_1 \beta_2  \\
 \alpha_2 \beta_1 - \alpha_1 \beta_3  \\
 \alpha_2 \beta_2 - \alpha_1 \beta_1  \\
\end{pmatrix} \\
\end{array} $} &
\multicolumn{3}{|c}{ $\begin{array}{l}
 \mathbf{3}\sim\begin{pmatrix}
 \alpha_2 \beta_3 - \alpha_1 \beta_2  \\
 \alpha_2 \beta_1 - \alpha_1 \beta_3  \\
 \alpha_2 \beta_2 - \alpha_1 \beta_1  \\
\end{pmatrix} \\
 \mathbf{3'}\sim\begin{pmatrix}
 \alpha_2 \beta_3 + \alpha_1 \beta_2  \\
 \alpha_2 \beta_1 + \alpha_1 \beta_3  \\
 \alpha_2 \beta_2 + \alpha_1 \beta_1  \\
\end{pmatrix} \\
\end{array} $} &
\multicolumn{3}{|c}{ $\begin{array}{l}
 \mathbf{\hat{3}}\sim\begin{pmatrix}
 \alpha_2 \beta_3 + \alpha_1 \beta_2  \\
 \alpha_2 \beta_1 + \alpha_1 \beta_3  \\
 \alpha_2 \beta_2 + \alpha_1 \beta_1  \\
\end{pmatrix} \\
 \mathbf{\hat{3}'}\sim\begin{pmatrix}
 \alpha_2 \beta_3 - \alpha_1 \beta_2  \\
 \alpha_2 \beta_1 - \alpha_1 \beta_3  \\
 \alpha_2 \beta_2 - \alpha_1 \beta_1  \\
\end{pmatrix} \\
\end{array} $} &
 \multicolumn{3}{|c|}{ $\begin{array}{l}
 \mathbf{\hat{3}}\sim\begin{pmatrix}
 \alpha_2 \beta_3 - \alpha_1 \beta_2  \\
 \alpha_2 \beta_1 - \alpha_1 \beta_3  \\
 \alpha_2 \beta_2 - \alpha_1 \beta_1  \\
\end{pmatrix} \\
 \mathbf{\hat{3}'}\sim\begin{pmatrix}
 \alpha_2 \beta_3 + \alpha_1 \beta_2  \\
 \alpha_2 \beta_1 + \alpha_1 \beta_3  \\
 \alpha_2 \beta_2 + \alpha_1 \beta_1  \\
\end{pmatrix} \\
\end{array} $} \rule[-8.5ex]{-4pt}{18ex} \\ \hline\hline

\multicolumn{6}{|c}{  $\mathbf{3} \otimes \mathbf{3} = \mathbf{3'} \otimes \mathbf{3'} = \mathbf{\hat{3}} \otimes \mathbf{\hat{3}} = \mathbf{\hat{3}'} \otimes \mathbf{\hat{3}'} = \mathbf{1} \oplus \mathbf{2} \oplus \mathbf{3} \oplus \mathbf{3'} $} & \multicolumn{6}{|c|}{ $\mathbf{3} \otimes \mathbf{3'} = \mathbf{\hat{3}} \otimes \mathbf{\hat{3}'} = \mathbf{1'} \oplus \mathbf{2} \oplus \mathbf{3} \oplus \mathbf{3'} $} \rule[-0.5ex]{-4pt}{3ex} \\ \hline
 \multicolumn{6}{|c}{ $\begin{array}{l}
 \mathbf{1}\sim \alpha_1 \beta_1+\alpha_2 \beta_3+\alpha_3 \beta_2 \\
 \mathbf{2}\sim\begin{pmatrix}
 \alpha_2 \beta_2+\alpha_1 \beta_3+\alpha_3 \beta_1 \\
 \alpha_3 \beta_3+\alpha_1 \beta_2+\alpha_2 \beta_1 \\
\end{pmatrix} \\
 \mathbf{3}\sim \begin{pmatrix}
 \alpha_3 \beta_2-\alpha_2 \beta_3 \\
 \alpha_2 \beta_1-\alpha_1 \beta_2 \\
 \alpha_1 \beta_3-\alpha_3 \beta_1 \\
\end{pmatrix}  \\
 \mathbf{3'}\sim \begin{pmatrix}
 2 \alpha_1 \beta_1-\alpha_2 \beta_3-\alpha_3 \beta_2 \\
 2 \alpha_3 \beta_3-\alpha_1 \beta_2-\alpha_2 \beta_1 \\
 2 \alpha_2 \beta_2-\alpha_1 \beta_3-\alpha_3 \beta_1 \\
\end{pmatrix} \\
  \end{array} $} &
 \multicolumn{6}{|c|}{ $  \begin{array}{l}
 \mathbf{1'}\sim \alpha_1 \beta_1+\alpha_2 \beta_3+\alpha_3 \beta_2\\
 \mathbf{2}\sim\begin{pmatrix}
 -(\alpha_2 \beta_2+\alpha_1 \beta_3+\alpha_3 \beta_1) \\
 \alpha_3 \beta_3+\alpha_1 \beta_2+\alpha_2 \beta_1 \\
\end{pmatrix} \\
 \mathbf{3}\sim\begin{pmatrix}
 2 \alpha_1 \beta_1-\alpha_2 \beta_3-\alpha_3 \beta_2 \\
 2 \alpha_3 \beta_3-\alpha_1 \beta_2-\alpha_2 \beta_1 \\
 2 \alpha_2 \beta_2-\alpha_1 \beta_3-\alpha_3 \beta_1 \\
\end{pmatrix} \\
 \mathbf{3'}\sim\begin{pmatrix}
 \alpha_3 \beta_2-\alpha_2 \beta_3 \\
 \alpha_2 \beta_1-\alpha_1 \beta_2 \\
 \alpha_1 \beta_3-\alpha_3 \beta_1 \\
\end{pmatrix} \\
\end{array} $} \rule[-12.5ex]{-4pt}{26ex} \\ \hline
 \multicolumn{6}{|c}{ $ \mathbf{3} \otimes \mathbf{\hat{3}} = \mathbf{3'} \otimes \mathbf{\hat{3}'} = \mathbf{\hat{1}} \oplus \mathbf{\hat{2}} \oplus \mathbf{\hat{3}} \oplus \mathbf{\hat{3}'} $} & \multicolumn{6}{|c|}{ $\mathbf{3} \otimes \mathbf{\hat{3}'} = \mathbf{3'} \otimes \mathbf{\hat{3}} = \mathbf{\hat{1}'} \oplus \mathbf{\hat{2}} \oplus \mathbf{\hat{3}} \oplus \mathbf{\hat{3}'} $} \rule[-0.5ex]{0pt}{3ex}\\ \hline
\multicolumn{6}{|c}{ $\begin{array}{l}
 \mathbf{\hat{1}}\sim \alpha_1 \beta_1+\alpha_2 \beta_3+\alpha_3 \beta_2 \\
 \mathbf{\hat{2}}\sim\begin{pmatrix}
 \alpha_2 \beta_2+\alpha_1 \beta_3+\alpha_3 \beta_1 \\
 \alpha_3 \beta_3+\alpha_1 \beta_2+\alpha_2 \beta_1 \\
\end{pmatrix} \\
 \mathbf{\hat{3}}\sim\begin{pmatrix}
 \alpha_3 \beta_2-\alpha_2 \beta_3 \\
 \alpha_2 \beta_1-\alpha_1 \beta_2 \\
 \alpha_1 \beta_3-\alpha_3 \beta_1 \\
\end{pmatrix} \\
 \mathbf{\hat{3}'}\sim\begin{pmatrix}
 2 \alpha_1 \beta_1-\alpha_2 \beta_3-\alpha_3 \beta_2 \\
 2 \alpha_3 \beta_3-\alpha_1 \beta_2-\alpha_2 \beta_1 \\
 2 \alpha_2 \beta_2-\alpha_1 \beta_3-\alpha_3 \beta_1 \\
\end{pmatrix} \\
\end{array} $} &
 \multicolumn{6}{|c|}{ $\begin{array}{l}
 \mathbf{\hat{1}'}\sim \alpha_1 \beta_1+\alpha_2 \beta_3+\alpha_3 \beta_2 \\
 \mathbf{\hat{2}}\sim\begin{pmatrix}
 -(\alpha_2 \beta_2+\alpha_1 \beta_3+\alpha_3 \beta_1) \\
 \alpha_3 \beta_3+\alpha_1 \beta_2+\alpha_2 \beta_1 \\
\end{pmatrix} \\
 \mathbf{\hat{3}}\sim\begin{pmatrix}
 2 \alpha_1 \beta_1-\alpha_2 \beta_3-\alpha_3 \beta_2 \\
 2 \alpha_3 \beta_3-\alpha_1 \beta_2-\alpha_2 \beta_1 \\
 2 \alpha_2 \beta_2-\alpha_1 \beta_3-\alpha_3 \beta_1 \\
\end{pmatrix} \\
 \mathbf{\hat{3}'}\sim\begin{pmatrix}
 \alpha_3 \beta_2-\alpha_2 \beta_3 \\
 \alpha_2 \beta_1-\alpha_1 \beta_2 \\
 \alpha_1 \beta_3-\alpha_3 \beta_1 \\
\end{pmatrix} \\
\end{array} $} \rule[-12.5ex]{-4pt}{26ex} \\ \hline\hline
\end{tabular} }
\caption{\label{tab:1}The CG coefficients of the $S_4\times Z_2$ Siegel modular group.}
\end{table}


\subsection{Modular forms with different weights for $\tau_1=\tau_2$ }
In Siegel modular form,
\begin{equation}
\tau= \left( \begin{matrix}
\tau_{1} & \tau_{3}\\
\tau_{3} & \tau_{2}
\end{matrix}\right)
\end{equation}

However, in our work, we strictly restrict our modular space for  $\tau_1=\tau_2$  such that there are four independent modular forms of weight 2 instead of five. These linearly independent modular forms of weight 2 can further be used using  Clebsch-Gordan coefficients as shown in Table~\ref{tab:1} to find linearly independent modular forms of weight $ k> 2$ in  $ S_4\times Z_2$ \cite{Ding:2020zxw}:
\begin{align}
\label{Y:S4xZ2}
\nonumber& \mathbf{3}':= ~~~Y_{\mathbf{3}'}(\tau)= \begin{pmatrix}
p_0(\tau)+4p_1(\tau)-p_3(\tau) \\
p_0(\tau)-2p_1(\tau)-p_3(\tau)-2i\sqrt{3}p_4(\tau) \\
p_0(\tau)-2p_1(\tau)-p_3(\tau)+2i\sqrt{3}p_4(\tau)
\end{pmatrix}\equiv\begin{pmatrix}
Y_1(\tau) \\ Y_2(\tau) \\ Y_3(\tau)
\end{pmatrix} \,,\\
&\mathbf{1}:= ~~~Y_{\mathbf{1}}(\tau)=p_0(\tau)+3p_3(\tau)\equiv Y_4(\tau)\,.
\end{align}

\begin{align}
\nonumber & \mathbf{1}:=~~~ \left\{
\begin{array}{l}
Y^{(4)}_{\mathbf{1}a} = Y_4^2\,,\\
Y^{(4)}_{\mathbf{1}b} = Y_1^2 + 2 Y_2Y_3\,,
\end{array}\right. \\
\nonumber&\mathbf{2}:= ~~~Y^{(4)}_{\mathbf{2}} = \begin{pmatrix}Y_2^2+2Y_1Y_3 \\ Y_3^2 +2Y_1Y_2 \end{pmatrix} \,, \\
\nonumber&\mathbf{3}:= ~~~Y^{(4)}_{\mathbf{3}} = (0,0,0)^T \,,\\
\label{Y4:S4xZ2} &\mathbf{3}':=~~~\left\{
\begin{array}{l}
Y^{(4)}_{\mathbf{3}'a} = Y_4
(Y_1 , Y_2 , Y_3 )^T \,,\\
Y^{(4)}_{\mathbf{3}'b} = 2 \begin{pmatrix}
Y_1^2-Y_2Y_3 \\ Y_3^2-Y_1Y_2 \\ Y_2^2-Y_1Y_3
\end{pmatrix}\,.
\end{array}
\right.
\end{align}

\begin{align}
\nonumber&\mathbf{1}:= ~~~
\left\{
\begin{array}{l}
Y^{(6)}_{\mathbf{1}a}= Y_4^3\,,\\
Y^{(6)}_{\mathbf{1}b}= Y_4Y_1^2 + 2 Y_2Y_3Y_4\,,\\
Y^{(6)}_{\mathbf{1}c} =2(Y_1^3+Y_2^3+Y_3^3-3Y_1Y_2Y_3)\,,\\
Y^{(6)}_{\mathbf{1}d} =Y_4Y_1^2 + 2 Y_2Y_3Y_4
\,,
\end{array}
\right.\\
\nonumber&\mathbf{2}:= ~~~
\left\{
\begin{array}{l}
Y^{(6)}_{\mathbf{2}a}= Y_4 \begin{pmatrix}
Y_2^2+2Y_1Y_3 \\ Y_3^2+2Y_1Y_2
\end{pmatrix} \,,\\
Y^{(6)}_{\mathbf{2}b}= (0,0)^T \,,\\
Y^{(6)}_{\mathbf{2}c}= Y_4 \begin{pmatrix}
Y_2^2+2Y_1Y_3 \\ Y_3^2+2Y_1Y_2
\end{pmatrix}\,,
\end{array}
\right.\\
\nonumber&\mathbf{3}:=~~~
\left\{
\begin{array}{l}
Y^{(6)}_{\mathbf{3}a}= 2 \begin{pmatrix}
Y_3^3-Y_2^3 \\ 2Y_1^2Y_2-Y_2^2Y_3-Y_3^2Y_1 \\ Y_2^2Y_1+Y_3^2Y_2-2Y_1^2Y_3
\end{pmatrix}  \,,\\
Y^{(6)}_{\mathbf{3}b}= (0,0,0)^T\,,\\
Y^{(6)}_{\mathbf{3}c}=  \begin{pmatrix}
Y_3^3-Y_2^3 \\ 2Y_1^2Y_2-Y_2^2Y_3-Y_3^2Y_1 \\ Y_2^2Y_1+Y_3^2Y_2-2Y_1^2Y_3
\end{pmatrix}  \,,
\end{array}
\right.
\end{align}

\begin{align}
\nonumber \mathbf{3}':= ~~~
\left\{
\begin{array}{l}
Y^{(6)}_{\mathbf{3}'a} = Y_4^2(Y_1, Y_2, Y_3)^T  \,,\\
Y^{(6)}_{\mathbf{3}'b}= 2 Y_4 \begin{pmatrix}
Y_1^2-Y_2Y_3 \\ Y_3^2-Y_1Y_2 \\ Y_2^2-Y_1Y_3
\end{pmatrix}   \,,\\
Y^{(6)}_{\mathbf{3}'c}= 2 \begin{pmatrix}
2Y_1^3- Y_2^3-Y_3^3 \\ 3Y_2^2Y_3-3Y_3^2Y_1 \\ 3Y_3^2Y_2-3Y_2^2Y_1
\end{pmatrix}   \,,\\
Y^{(6)}_{\mathbf{3}'d}=  \begin{pmatrix}
Y_2^3+Y_3^3+ 4Y_1Y_2Y_3 \\ 3Y_3^2Y_1+2Y_1^2Y_2 +Y_2^2Y_3 \\ 3Y_2^2Y_1+2Y_1^2Y_3+Y_3^2Y_2
\end{pmatrix}   \,,\\
Y^{(6)}_{\mathbf{3}'e}= 2 Y_4 \begin{pmatrix}
Y_1^2-Y_2Y_3 \\ Y_3^2-Y_1Y_2 \\ Y_2^2-Y_1Y_3
\end{pmatrix}   \,,\\
Y^{(6)}_{\mathbf{3}'f}= Y_4^2(Y_1, Y_2, Y_3)^T \,,\\
Y^{(6)}_{\mathbf{3}'g}= (Y_1^2+2Y_2Y_3)(Y_1, Y_2, Y_3)^T \,,
\end{array}
\right.
\end{align}

Using Fourier expansion, we expand $Y_{1,2,3,4}$ from Eqns~\ref{Y:S4xZ2} as \cite{Ding:2020zxw}:
\begin{align}
\nonumber
Y_1(\tau)&=1+32q_1^{\frac{1}{2}}-q_1(8q_3^{-1}+8q_3)+q_1^{\frac{3}{2}}(512+192q_3^{-1}+192q_3) +q_1^2(64+24q_3^{-2}+24q_3^2)\\
\nonumber&+q_1^{\frac{5}{2}}(1152+416q_3^{-2}+1024q_3^{-1}+1024q_3+416q_3^2)+ q_1^{3}(-32 q_3^{-3}-192q_3^{-1}+192q_3-32q_3^3) \\
\nonumber&+q_1^{\frac{7}{2}}(2048+448q_3^{-3}+1536q_3^{-2}+2496q_3^{-1}+2496q_3+1536q_3^2+448q_3^3)+\dots\\
\nonumber
Y_2(\tau)&=1-16q_1^{\frac{1}{2}}-q_1(8q_3^{-1}+64i\sqrt{3}q_3^{-\frac{1}{2}}+64i\sqrt{3}q_3^{\frac{1}{2}}+8q_3)-q_1^{\frac{3}{2}}(256+96q_3^{-1}+96q_3)+q_1^{2}(64 \\
\nonumber&+ 24q_3^{-2}-128i\sqrt{3}q_3^{-\frac{3}{2}} -384i\sqrt{3}q_3^{-\frac{1}{2}}-384i\sqrt{3}q_3^{\frac{1}{2}}-128i\sqrt{3}q_3^{\frac{3}{2}}+24q_3^2)-q_1^{\frac{5}{2}}(576 +208q_3^{-2}\\
\nonumber& +512q_3^{-1}+512q_3+208q_3^2) - q_1^{\frac{7}{2}}(1024+224q_3^{-3}+768q_3^{-2}+1248q_3^{-1}+1248q_3+768q_3^2\\
\nonumber&+224q_3^3)+\dots \\
\nonumber
Y_3(\tau)&=1-16q_1^{\frac{1}{2}}-q_1(8q_3^{-1}-64i\sqrt{3}q_3^{-\frac{1}{2}}-64i\sqrt{3}q_3^{\frac{1}{2}}+8q_3)-q_1^{\frac{3}{2}}(256+96q_3^{-1}+96q_3)+q_1^{2}(64 \\
\nonumber&+ 24q_3^{-2}+128i\sqrt{3}q_3^{-\frac{3}{2}} +384i\sqrt{3}q_3^{-\frac{1}{2}}+384i\sqrt{3}q_3^{\frac{1}{2}}+128i\sqrt{3}q_3^{\frac{3}{2}}+24q_3^2)-q_1^{\frac{5}{2}}(576 +208q_3^{-2}\\
\nonumber& +512q_3^{-1}+512q_3+208q_3^2) - q_1^{\frac{7}{2}}(1024+224q_3^{-3}+768q_3^{-2}+1248q_3^{-1}+1248q_3+768q_3^2\\
\nonumber &+224q_3^3)+\dots\,,\\
\nonumber
Y_4(\tau)&=1+q_1(192+24q_3^{-1}+24q_3)+q_1^2(576+24q_3^{-2}+768q_3^{-1}+768q_3+24q_3^2)+q_1^3(3072 \\
\nonumber&+96q_3^{-3}+1152q_3^{-2}+576q_3^{-1}+576q_3 +1152q_3^2+96q_3^3) + q_1^4(576+24 q_3^{-4}+1536q_3^{-3} \\
&+2304q_3^{-2}+4608q_3^{-1}+4608q_3+2304q_3^2+1536q_3^3+24q_3^4)+\dots \,,
\end{align}
where $q_1=e^{2\pi i\tau_1}$ and $q_3=e^{2\pi i\tau_3}$.
\end{appendices}

\providecommand{\href}[2]{#2}\begingroup\raggedright
\endgroup

\end{document}